\documentclass[reprint,superscriptaddress,amsmath,amssymb,aps,pra,floatfix]{revtex4-2}
\usepackage[utf8]{inputenc}
\usepackage{array}
\usepackage{graphicx}
\usepackage{dcolumn}
\usepackage{bm}
\usepackage{xcolor}
\usepackage{colortbl}
\usepackage{ulem} 
\usepackage{soul}
\usepackage{mathrsfs}
\usepackage{hyperref} 
\usepackage{cleveref} 

\hypersetup{colorlinks = true, 
	    linkcolor = blue, 
	    urlcolor = blue,
            citecolor = blue} 

\usepackage{setspace} 

\begin{document}
\begin{spacing}{1.0}
\title{Efficient biphoton generation by a waveguide-coupled single atom}
\author{Mao-Hua Wang}
\affiliation{School of Physics and Center for Quantum Sciences, Northeast Normal University, Changchun 130024, China}
\author{Xiao-Jun Zhang}
\affiliation{School of Physics and Center for Quantum Sciences, Northeast Normal University, Changchun 130024, China}
\author{M. Artoni}
\email{artoni@lens.unifi.it}
\affiliation{Department of Engineering and Information Technology, Brescia University, 25133 Brescia, Italy}
\affiliation{European Laboratory for Nonlinear Spectroscopy (LENS), 50019 Sesto Fiorentino, Italy}
\author{G. C. La Rocca}
\email{g.larocca@sns.it}
\affiliation{NEST, Scuola Normale Superiore, 56126 Pisa, Italy}
\author{Jin-Hui Wu}
\email{jhwu@nenu.edu.cn}
\affiliation{School of Physics and Center for Quantum Sciences, Northeast Normal University, Changchun 130024, China} 
\date{\today}

\begin{abstract}
A single atom undergoing spontaneous four-wave mixing near a chiral waveguide can efficiently channel an emitted Stokes-anti-Stokes photon pair into two tightly confined waveguide modes, yielding thus enhanced biphoton generation without requiring loss suppression or stringent phase matching. We develop a perturbative treatment, valid for a four-level atomic system under experimentally realistic conditions, to explain physical origins and clarify relevant constraints of such an enhancement determined by the interplay of atomic decay rates toward guided and unguided modes. Besides achieving optimal generation rates equivalent to a cold atomic ensemble hundreds of micrometers long in free space, our biphoton source naturally fulfills key requirements for next-generation on-chip quantum light sources, namely low-loss operation, robustness, compactness, and scalability.
\end{abstract}
\maketitle

\textit{Introduction.} - Developing high-performance quantum light sources is pivotal for the implementation of quantum communication, computation, metrology, etc. Generating photon pairs or \textit{biphotons} via nonlinear processes in free space remains one of the most established paradigms. Most current implementation schemes employ spontaneous parametric down-conversion (SPDC) in nonlinear crystals~\cite{SPDC1,SPDC2} or spontaneous four-wave mixing (SFWM) in atomic ensembles~\cite{SFWM1,SFWM2,SFWM3,SFWM4,SFWM5,SFWM6,SFWM7,SFWM8}. The generation rates of narrowband biphotons in driven atoms usually reach the range of $\{10^3,10^4\}$ $s^{-1}$ under realistic conditions~\cite{rate1,rate2,rate3,rate4,rate5,rate6,rate7}, providing a robust platform for investigating various quantum phenomena among which entanglement may be the most ubiquitous one. Despite these advances, conventional free-space solutions suffer from several intrinsic bottlenecks. First, experimental setups are generally bulky and structurally complex, severely restricting  scalability to chip-scale integration~\cite{cs1}. Second, photonic interactions with a large number of atoms in random motion inevitably introduce parasitic noises and losses, thereby compromising the overall biphoton generation performance~\cite{loss}. More importantly, biphotons probabilistically emitted into free space exhibit significant spatial divergence or poor spatial mode resolution, thereby leading to one to two-order of magnitude reduction in the available rates when coupled to single-mode nanofibers or any integrated photonic devices. This  poses significant challenges for practical applications~\cite{qp1}.

Recent advances in waveguide quantum electrodynamics (QED)~\cite{QED1,QED2,QED3,QED4,QED5,QED6,QED7,QED9,QED10,QED11,QED12,QED13,QED14,QED15,QED16,QED17,QED18}, on the other hand, have unveiled a wealth of remarkable results ranging from topologically protected chiral bound states~\cite{lr1,lr2} and collective effects such as super and subradiance~\cite{su1}  to efficient single-photon frequency conversion~\cite{fc1,fc2} and photon-mediated interactions between quantum emitters~\cite{dot1,dot2,dot3}.
A growing number of these results underscore the emergence of waveguide QED as a rapidly evolving frontier of quantum optics providing a powerful (conceptual) tool for engineering light-matter interactions between transversely confined photons and individual quantum emitters, whether in the form of real or artificial atoms. A waveguide coupled to a ring cavity of second-order nonlinearity has been proposed to generate biphoton states via parametric decay of single photons but bears a limited efficiency due to intrinsic low nonlinearity-to-loss ratios~\cite{ring}. In addition, the possibility of controlling light-matter interaction in chiral waveguides has garnered extensive attention~\cite{loda,hafe} by offering fundamentally new functionalities and applications. Atom and giant-atom based chiral waveguide QED studies have predominantly been restricted to photon scattering within the linear-response regime~\cite{sc1,sc2} or, very recently, to nonlinear photon conversion~\cite{our}. Genuinely quantum nonlinear effects mediated by quantum emitters coupled to confined waveguide platforms remain largely unexplored yet.

In the effort to bridge this gap, we here show that the nonlinear SFWM generation of high-quality biphotons become particularly efficient in a chiral waveguide coupled to a single four-level atom driven by external (far-detuned) pump and (resonant) coupling fields. Prominent generation rates and significant nonclassical features observed at the Stokes and anti-Stokes pair of waveguide modes stem from tight transverse confinement, which allows us to control atom-photon coupling strengths hence tailor atomic decay rates into the two waveguide modes. Such an enhanced generation performance mainly constrained by low atomic excitation and moderate biphoton probability are largely determined by the interplay between these decay rates and that into unguided radiation modes. This is supported by a perturbative treatment developed in the (experimentally feasible) weak-to-moderate atom-waveguide coupling regime with a proper truncation of the atom-photon state hierarchy.

\begin{figure}[ptbh]
\includegraphics[width=8.5 cm]{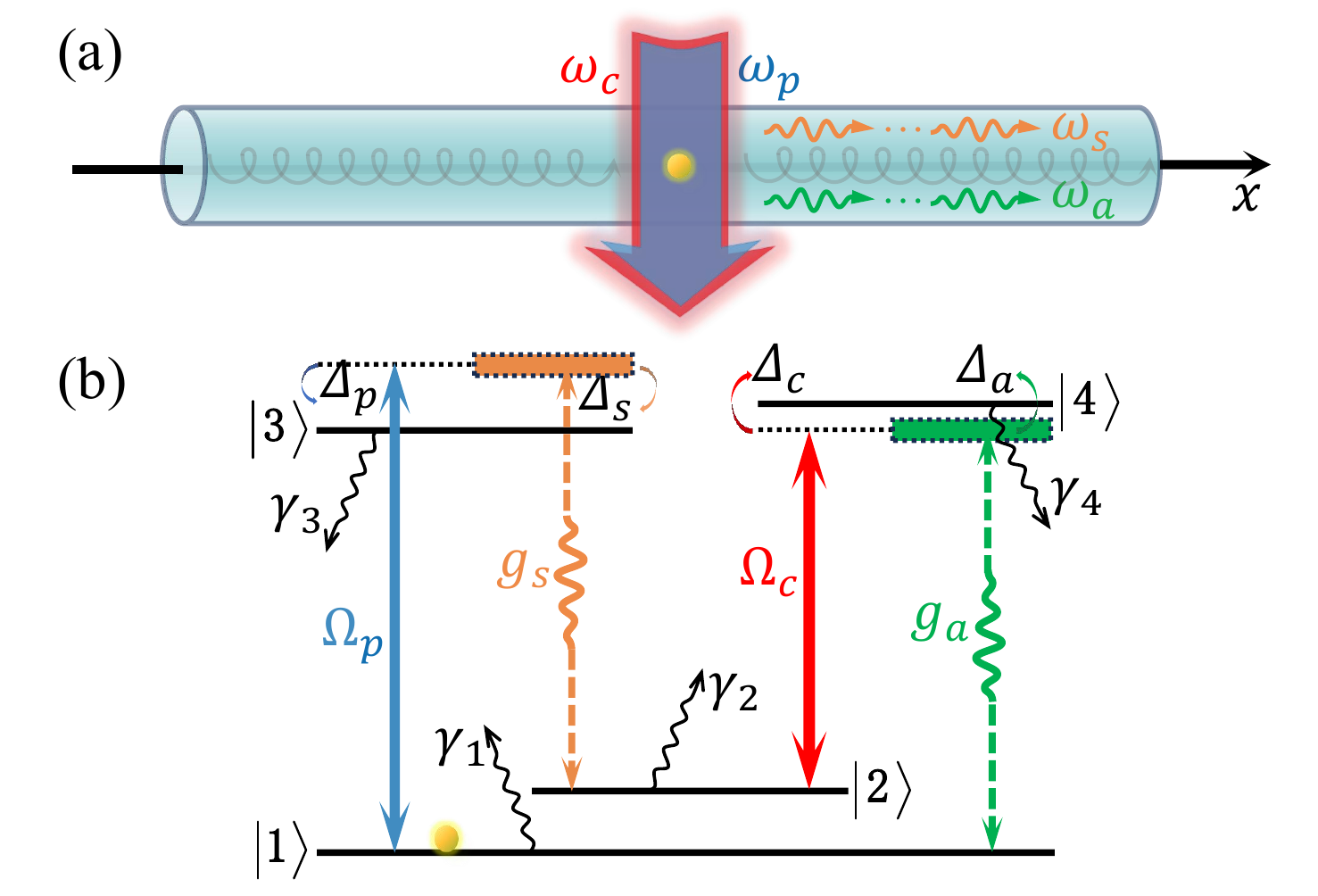}
\caption{(a) Schematic illustration of the paired generation of a Stokes ($\omega_{s}$) and an anti-Stokes ($\omega_{a}$) photon from a single atom trapped close to a chiral helical waveguide when illuminated by a pump ($\omega_{p}$) and a coupling ($\omega_{c}$) field. (b) The double-$\Lambda$ configuration of a four-level atom allowing for the closed-loop SFWM process $|1\rangle \to |3\rangle \to |2\rangle \to |4\rangle \to |1\rangle$ of amplitude decay rates $\gamma_{1}$, $\gamma_{2}$, $\gamma_{3}$, and $\gamma_{4}$ toward free space. The pump and coupling fields act upon transitions $|1\rangle \leftrightarrow |3\rangle$ and $|2\rangle \leftrightarrow |4\rangle$ with Rabi frequencies (detunings) $\Omega_{p}=\mu_{31}E_{p}/2\hbar$ and $\Omega_{c}=\mu_{42}E_{c}/2\hbar$ ($\Delta_{p}=\omega_{p}-\omega_{31}$ and $\Delta_{c}=\omega_{c}-\omega_{42}$), respectively. Continuous chiral waveguide modes near resonant with transitions $|2\rangle \leftrightarrow |3\rangle$ and $|1\rangle \leftrightarrow |4\rangle$ facilitate the directional generation of Stokes and anti-Stokes photons with coupling strengths (detunings) $g_{s}=\mu_{32}\mathcal{E}_{s}/2\hbar$ and $g_{a}=\mu_{41}\mathcal{E}_{a}/2\hbar$ ($\Delta_{s}=\omega_{s}-\omega_{32}$ and $\Delta_{a}=\omega_{a}- \omega_{41}$), respectively.}\label{fig1}
\end{figure}

Optimal generation rates over $10^{4}$ s$^{-1}$ can be attained for an \textit{alkali-metal} atom with the concomitant suppression of single-photon and multiple-biphoton events, leading to reduced noise generation and an enhanced photon-pairing probability. While these rates comparable to that in cold atomic samples of density $10^{11}\,cm^{-3}$ and length $100 \div 600$ $\mu m$~\cite{rate1,rate2,rate3,rate4,rate5,rate6,rate7}, our biphoton source offers a further practical advantage by mitigating the substantial losses associated with mode-mismatched coupling into optical fibers, that is a critical issue in quantum networking operations. Replacing the optical nonlinearity of a bulky free-space atomic ensemble with that of a single-atom trapped near a versatile chiral waveguide, where photon propagation can be readily tailored through dispersion engineering~\cite{dis1}, naturally eases direct integration into a nanoscale photonic-chip type of platform~\cite{ch1,ch2,ch3}. The approach we present is thus meant to establish a sound physical ground for the development of (next-generation) on-chip quantum light sources that combine enhanced generation rates and low-loss operation with robustness and compactness for high-scalability. 

\textit{Model and Equations.} - We show in Fig.~\ref{fig1} a prototype platform for realizing the waveguide-enhancement of a SFWM process mediated by a single four-level atom via the closed-loop (double-$\Lambda$) transition path $|1\rangle \to |3\rangle \to |2\rangle \to |4\rangle \to |1\rangle$. When illuminated by classical pump (of frequency $\omega_{p}$) and coupling (of frequency $\omega_{c}$) beams, the atom trapped in proximity of the waveguide may spontaneously emit a pair of Stokes and anti-Stokes photons at frequencies $\omega_{s}$ and $\omega_{a}$, respectively. For a suitable choice of atomic states, both photons can be taken left-circularly polarized (LCP) and moving to the right within the chiral waveguide. Using a far-detuned pump we can make the subsystem $|1\rangle\leftrightarrow |3\rangle \leftrightarrow |2\rangle$ act as a Raman transition path for the Stokes photon. Conversely, under resonant coupling, the subsystem $|1\rangle\leftrightarrow |4\rangle \leftrightarrow |2\rangle$ induces electromagnetically induced transparency~(EIT)~\cite{SFWM1,SFWM2,SFWM3,SFWM4,SFWM5,SFWM6,SFWM7,SFWM8} for the anti-Stokes photon. While atomic population is mostly confined to ground state $|1\rangle$, such a pump and coupling choice suppresses the linear single-atom response and simultaneously enhancing its nonlinear response, thereby increasing the desired SFWM nonlinearity.

Based on the above considerations and assuming that the nonlinear response remains sufficiently weak to avoid generating states containing more than one Stokes photon and one anti-Stokes photon, we develop a perturbative treatment of the SFWM process in which the pump and coupling (classical) fields are kept to all orders, while the Stokes and anti-Stokes (quantum) fields are kept only to the lowest order.
The smallest relevant manifold of involved states comprises a state with a single Stokes photon and a state containing one Stokes and one anti-Stokes photon, besides the vacuum (state) with  neither Stokes nor anti-Stokes photons. This truncation of the state hierarchy excludes contributions from processes generating higher photon numbers. The dynamics of this single-atom two-mode-waveguide hybrid system can then be described by
\begin{widetext}
\begin{equation}\label{Eq1}
\vert\psi(t)\rangle=\left[u_{1}(t)\vert1\rangle+u_{3}(t)\vert3\rangle\right]\vert0_s,0_a\rangle+
\int d\omega_s\left[u_{2}^s(t)\vert2\rangle +u_{4}^s(t)\vert4\rangle\right]\vert1_s,0_a\rangle
+\int\int d\omega_s d\omega_a\left[u_{1}^{sa}(t)\vert1\rangle +u_{3}^{sa}(t)\vert3\rangle\right]\vert1_s,1_a\rangle,
\end{equation}
\end{widetext}
which extends the single-excitation ansatz for spontaneous emission, in which either an atom or a waveguide mode is excited, to the regime where the atom and two waveguide modes could be simultaneously excited. As a result, the Hilbert space must encompass parts of the two- and three-excitation manifolds. Here $u_1(t)$ and $u_3(t)$ refer to the probability amplitudes of finding the atom respectively in states $|1\rangle$ and $|3\rangle$, with no Stokes or anti-Stokes photon ($\vert 0_s,0_a \rangle$) being excited; $u_{2}^{s}(t)$ and $u_{4}^{s}(t)$ represent the probability amplitude densities for generating a single Stokes photon  ($\vert 1_s,0_a \rangle$) at frequency $\omega_s$ with the atom making a two-photon transition to state $|2\rangle$ after absorbing a pump photon or a three-photon transition to state $|4\rangle$ after absorbing an additional coupling photon; $u_{1}^{sa}(t)$ and $u_{3}^{sa}(t)$ describe the joint probability amplitude densities for generating a Stokes and anti-Stokes photon pair ($\vert 1_s,1_a \rangle$) at frequencies $\omega_s$ and $\omega_a$, occurring either when the atom returns to state $|1\rangle$ completing the SFWM process or when the atom is excited again to state $|3\rangle$ involving the absorption of another pump photon.

The dynamics, which continuously transfer probability amplitudes among the four atomic states in a closed-loop path through populations of the two-excitation manifold of waveguide photons ($\{\vert 0_s,0_a \rangle, \vert 1_s,0_a \rangle, \vert 1_s,1_a \rangle\}$), is governed by the following Hamiltonian in the electric-dipole and rotating-wave approximations
\begin{widetext}
\begin{equation}\label{Eq2}
\begin{split}
\frac{\mathcal{H}_{tot}}{\hbar}=&-(i\gamma_{1}\vert1\rangle\langle1\vert +\delta_{p}^{*}\vert3\rangle\langle3\vert)
\vert0_{s},0_{a}\rangle|\langle0_{s},0_{a}\vert +\int d\omega_{s}(\delta_{sp}\vert2\rangle\langle2\vert+\delta_{cp}\vert4\rangle\langle4\vert)
\vert1_{s},0_{a}\rangle\langle1_{s},0_{a}\vert +\int\int d\omega_{s}d\omega_{a}\\
&[\delta_{ap}\vert1\rangle\langle1-(\delta_{p}^{*}+i\Gamma_{s})\vert3\rangle\langle3]\vert\vert 1_{s},1_{a}\rangle\langle 1_{s},1_{a}\vert-\big[\Omega_{p}\vert3\rangle\langle1\vert\vert 0_{a},0_{a}\rangle\langle0_{s},0_{a}\vert +\int d\omega_{s}(\Omega_{c}|4\rangle\langle 2||1_{s},0_{a} \rangle\\ 
&+\tilde{g}_{s}|3\rangle\langle2||0_s,0_a \rangle)\langle 1_s,0_a|+\int\int d\omega_sd\omega_{a}(\Omega_{p}\vert 3\rangle\langle 1\vert \vert 1_s,1_a \rangle + \tilde{g}_a \vert 4\rangle\langle 1\vert
\vert 1_s,0_a \rangle)\langle 1_s,1_a\vert +h.c.\big],\\
\end{split}
\end{equation}
\end{widetext}
where we introduced the complex detunings $\delta_{p}=\Delta_{p}-i\gamma_{3}$, $\delta_{sp}=\Delta_{sp}-i\gamma_{2}=\Delta_{s}-\Delta_{p}-i\gamma_{2}$,
$\delta_{cp}=\Delta_{cp}-i\gamma_{4}=\Delta_{sp}-\Delta_{c}-i\gamma_{4}$, and $\delta_{ap}=\Delta_{ap}-i\gamma_{1}=\Delta_{cp}+\Delta_{a}-i\gamma_{1}$ in terms of $\Delta_{p,s,c,a}$ and $\gamma_{1,2,3,4}$ (see Fig.~\ref{fig1}). We have also introduced an extra decay rate $\Gamma_{s}$ (as defined later)
of atomic level $|3\rangle$ to account for its coupling to truncated multiple Stokes and anti-Stokes biphotons. The single-photon excitations $|1_{\alpha}\rangle=a^{\dagger}_{\alpha}|0_{\alpha}\rangle$ are described by the continuous-mode representation of (creation) operators $a_\alpha^\dagger$ for $\alpha=\{s,a\}$ satisfying the familiar commutation relation $[a_\alpha, a_{\alpha'}^\dagger] = \delta(\omega_\alpha -\omega_{\alpha'})$. Consistency with these operators (units of $s^{1/2}$) requires a rescaling of the coupling strengths $g_\alpha$ by the coherence time $\tau_{\alpha}=L_{\alpha}/v_{g}$, \textit{i.e.} $\tilde{g}_{\alpha}=g_{\alpha}\sqrt{\tau_{\alpha}}=(\mu_{\alpha}\mathcal{E}_{\alpha}/2\hbar)\sqrt{L_{\alpha}/v_{g}}=(\mu_{\alpha}/2)\sqrt{\omega_{\alpha}/\hbar\pi\epsilon_{0}v_{g}A}$
where $L_{\alpha}$, $v_{g}$, and $A$ denote, respectively, the photon coherence length, group velocity, and cross-sectional area, the last of which can strongly influence the photon-pair generation rate of our hybrid system as examined below. Also, to ensure dimensional consistency with continuous-mode operators $a_\alpha^\dagger$, $u_{1,3}(t)$ must be dimensionless, $u_{2,4}^{s}(t)$ are in units of $s^{1/2}$, whereas $u_{1,3}^{sa}(t)$ are in units of $s$.

Substituting $\mathcal{H}_{tot}$~in Eq.~(\ref{Eq2}) and $|\psi(t)\rangle$ in Eq.~(\ref{Eq1}) into the Schrödinger equation $i\hbar \partial_t |\psi\rangle = \mathcal{H}_{tot}|\psi\rangle$, we obtain a set of dynamic equations (regarding $u_{1,3}$, $u_{2,4}^{s}$, and $u_{1,3}^{sa}$) whose steady-state solutions of our interest read as
\begin{equation}\label{Eq3}
\begin{split}
\frac{u_{2}^{s}}{u_{1}}&=\frac{-(\Delta_{sp}-i\gamma_{4a})\Omega_p\tilde{g}_s^*}{(\Delta_p+i\gamma_{3s})(\Delta_{sp}-\Omega_{e}-i\gamma_{e})(\Delta_{sp}+\Omega_{e}-i\gamma_{e})},\\
\frac{u_{4}^{s}}{u_{1}}&=\frac{-\Omega_p \tilde{g}_s^* \Omega_c}{(\Delta_p+i\gamma_{3s})(\Delta_{sp}-\Omega_{e}-i\gamma_{e})(\Delta_{sp}+\Omega_{e}-i\gamma_{e})},\\
\frac{u_{1}^{sa}}{u_{1}}&=\frac{-\Omega_p \tilde{g}_s^* \Omega_c \tilde{g}_a^*/(\Delta_{ap}^{\prime}-i\gamma_{1}^{\prime})}{(\Delta_p+i\gamma_{3s})(\Delta_{sp}-\Omega_{e}-i\gamma_{e})(\Delta_{sp}+\Omega_{e}-i\gamma_{e})},
\end{split}
\end{equation}
with $\Omega_e\simeq \sqrt{|\Omega_c|^2-\gamma_e^2}$, $\gamma_e\simeq\gamma_{4a}/2$, $\gamma_{3s}=\gamma_{3}+\Gamma_{s}$, and $\gamma_{4a}=\gamma_{4}+\Gamma_{a}$. Here, $\Gamma_{s,a}=\pi|\tilde{g}_{s,a}|^2$ are the decay rates of atomic levels $|3\rangle$ and $|4\rangle$ into Stokes and anti-Stokes modes, \textit{viz.} the fraction of spontaneous emission redirected into \textit{guided} modes at variance with $\gamma_{3,4}$ that represent emission into \textit{unguided} radiation modes. These hold only for $\Delta_c = 0$, $\gamma_{1,2} \to 0$, and  $|\Delta_{p}+i\gamma_{3s}|\gg\Gamma_{s}$ while the general results and their derivation are discussed below. As a result, both single-photon and biphoton overall probability amplitude densities,
\begin{equation}\label{Eq4}
\begin{split}
|u_{2}^s|^2+|u_{4}^s|^2&=|\tilde g_{s}|^2 S_{p} \frac{(\Delta_{sp}^2+\gamma_{4a}^2+|\Omega_c|^2)}{D_e}|u_{1}^2|,\\
|u_{1}^{sa}|^2&=|\tilde g_s|^2 S_{p}\frac{|\Omega_c|^2|\tilde{g}_a|^2}{D_{e}(\Delta_{ap}^{\prime2}+\gamma_{1}^{\prime2})}|u_{1}|^2,
\end{split}
\end{equation}
are directly proportional to $S_{p}\simeq|\Omega_p|^2\,/(\Delta_p^2+\gamma_{3s}^2)$, the pump-transition saturation parameter, which must remain small for ensuring the validity of our model, \textit{i.e.} a low steady-state population ($S_{p}|u_{1}|^2\ll1$) of atomic level $|3\rangle$. The spectral characteristics of $|u_{2}^s|^2+|u_{4}^s|^2$ is governed by 
$D_e=\left[(\Delta_{sp}-\Omega_e)^2+\gamma_e^2\right]\left[(\Delta_{sp}+\Omega_e)^2+\gamma_e^2\right]$, a pole denominator with effective Rabi frequency $\Omega_{e}$ determining two-photon symmetric resonances located at $\Delta_{sp}=\pm\Omega_e$ whereas effective dephasing rate $\gamma_e$ setting the common linewidth of these resonances. 

Similarly for the biphoton probability amplitude density $|u_{1}^{sa}|^2$, which is governed as before by $D_{e}$ but also by the modified detuning
$\Delta_{ap}^{\prime}=\Delta_{ap}+\Delta_{p}S_p$ and dephasing $\gamma_{1}^{\prime}=\gamma_{1}+\gamma_{3s}S_{p}$. Its pole $\Delta_{ap}^{\prime}=0$ is a manifestation of exact energy conservation in the SFWM process and turns out to be $\Delta_{ap}=0$ within the usual single-mode treatment. A finite SFWM linewidth $\gamma_{1}^{\prime}$, equaling the pump-transition scattering rate $\gamma_{3s}S_p$ due to $\gamma_{1}\to0$, is clearly necessary to remove the pole singularity. Within this context, it is also worth recalling that the contribution of $|u_{3}^{sa}|^2$ to biphoton generation, arising from the second-order correction $|3\rangle |1_s,1_a\rangle$ in $|\psi(t)\rangle$, carries a negligible weight to an excited state off a predominantly populated ground state and it has been excluded from Eq.~(\ref{Eq4}). In particular, one has $|u_{3}/u_{1}|^2\simeq|u_{3}^{sa}|^2/|u_{1}^{sa}|^2\simeq S_p$, much less than unity for our perturbative approach to hold and provided $\Gamma_{s,a}<\gamma_{3,4}$. This last inequality indicates that the associated beta factors $\beta = \Gamma/(\Gamma+\gamma)$ fall into the weak-to-moderate coupling regime rather than the strong-coupling regime so that this scheme becomes significantly more feasible for practical realization~\cite{QED7}.

With the solutions in Eq.~(\ref{Eq4}) at hand, the probability of generating a single Stokes photon, $P_s=\int[|u_{2}^s|^2+|u_{4}^s|^2]d\omega_s$, and that of generating a Stokes–anti-Stokes photon pair, $P_{sa}=\int\int|u_{1}^{sa}|^2d\omega_{s}d\omega_{a}$, can be evaluated analytically by means of contour integration. For small values of $S_p$, taking into account that $P_{sa}+|u_{1}|^{2}\to 1$, we obtain
\begin{equation}\label{Eq5}
\begin{split}
P_{s}&=\frac{2S_{p}\Gamma_s|u_{1}|^{2}}{\gamma_{4a}}\simeq\frac{2S_{p}\Gamma_s\gamma_{3s}}{\gamma_{3s}\gamma_{4a}+\Gamma_{s}\Gamma_{a}},\\
P_{sa}&=\frac{\Gamma_s\Gamma_a|u_{1}|^{2}}{\gamma_{3s}\gamma_{4a}}\simeq
\frac{\Gamma_{s}\Gamma_{a}}{\gamma_{3s}\gamma_{4a}+\Gamma_{s}\Gamma_{a}},
\end{split}
\end{equation}
which clearly shows that, in the regime of interest \textit{viz.} $|\Omega_{p}|^{2}\ll(\Delta_{p}^{2}+\gamma_{3s}^{2})$, one has $P_{sa}\gg P_{s}\to 0$ to guarantee a high purity of the SFWM pair generation process.

For a Stokes-anti-Stokes photon pair generated via a SFWM process, its second-order intensity correlation can be defined in terms of the time-dependent photon operators $b_{\alpha}(t_{\alpha})$ as a Fourier transform of $a_{\alpha}(\omega_{\alpha})$ by
\begin{equation}\label{Eq6}
\begin{split}
G_{sa}^{(2)}(t_s,t_a)&=|\langle 1,0_s,0_a \vert b_s(t_s)b_a(t_a)|\psi\rangle|^2\\
&\simeq \big|\frac{1}{2\pi}\int\int d\omega_s d\omega_a e^{-i\omega_s t_s}e^{-i\omega_a t_a} u_{1}^{sa}\big|^2,
\end{split}
\end{equation}
where in the last expression the tiny contribution of $u_{3}^{sa}$ is neglected. Applying the residue theorem to evaluate the two-fold frequency integrals involved in Eq.~(\ref{Eq6}), it is easy to derive a compact analytical expression
\begin{equation}\label{Eq7}
G_{sa}^{(2)}(\tau)\simeq
\frac{4S_{p}\Gamma_{s}\Gamma_{a}\gamma_{3s}\gamma_{4a}}{(\gamma_{3s}\gamma_{4a}+\Gamma_s \Gamma_a)}e^{-\gamma_{4a}\tau}\sin^2(\Omega_{e}\tau),
\end{equation}
as a function of time delay $\tau=t_{a}-t_{s}$. Then, the biphoton generation rate can be calculated as
\begin{equation}\label{Eq8}
R_{sa}=\int d\tau\; G_{sa}^{(2)}(\tau)\simeq\frac{2S_{p}\Gamma_s\Gamma_a\gamma_{3s}}{(\gamma_{3s}\gamma_{4a}+\Gamma_{s}\Gamma_{a})}=2 \gamma_1^{\prime}\,P_{sa},
\end{equation}
where the last expression shows that in the steady-state regime the biphoton generation rate balances the depletion rate of the biphoton population in state $|\psi_1^{sa}\rangle$. The rate $R_{sa}$ should be made as large as possible by increasing $\Gamma_{s,a}$, restricted \textit{e.g.} by $P_{sa}<1/10$ for $S_{p}\simeq1/100$, from the perspective of practical applications. It can also be used to calculate the normalized cross-correlation function $g_{sa}^{(2)}(\tau)=G_{sa}^{(2)}(\tau)/R_{sa}^{2}$ quantifying the nonclassical feature of a Stokes and anti-Stokes biphoton. Moreover, above approximate expressions indicate a spectrum-insensitive biphoton generation because $P_{s}$, $P_{sa}$, and $R_{sa}$ are independent of $\Omega_{c}$ determining the spectral structures of $u_{2}^{s}$, $u_{4}^{s}$, and $u_{1}^{sa}$.

\begin{figure}[ptbh]
\includegraphics[width=8.5 cm]{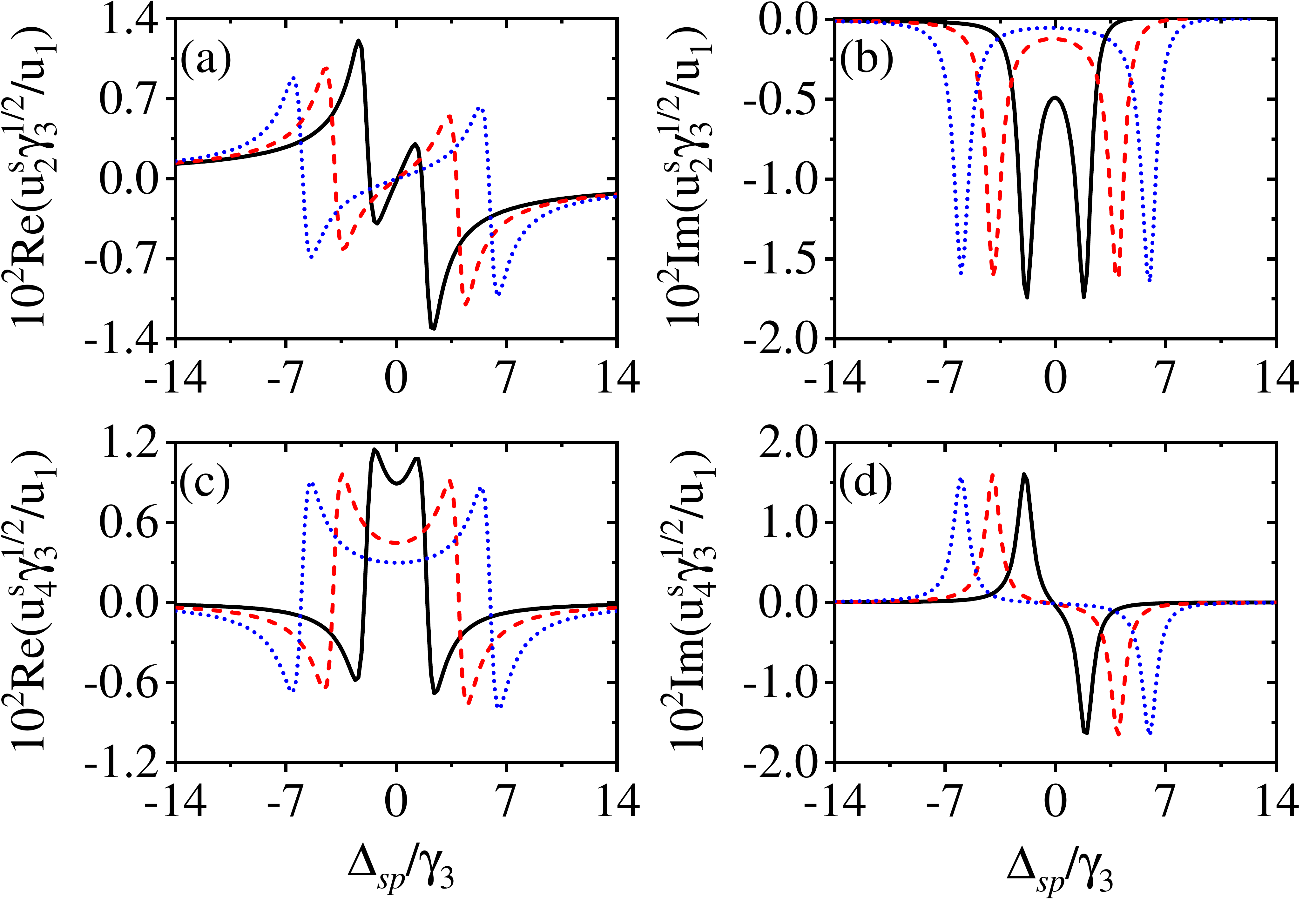}
\caption{Real (a) and imaginary (b) parts of $10^{2}u_{2}^{s}/u_{1}$ in units of $\gamma_{3}^{-1/2}$ against $\Delta_{sp}/\gamma_{3}$ or real (c) and imaginary (d) parts of $10^{2}u_{4}^{s}/u_{1}$ in units of $\gamma_{3}^{-1/2}$ against $\Delta_{sp}/\gamma_{3}$. Black-solid, red-dashed, and blue-dotted lines refer to $\Omega_{c}/\gamma_{3}=2.0$, $4.0$, and $6.0$, respectively. Other parameters are $\Omega_{p}/\gamma_{3}=1.0$, $\Delta_{p}/\gamma_{3}=10$, $\Delta_{c}=0$, $\Gamma_{s}/\gamma_{3}=\Gamma_{a}/\gamma_{3}=0.1$, and $\gamma_{4}=\gamma_{3}$.}\label{fig2} 
\end{figure}

\begin{figure}[ptbh]
\includegraphics[width=8.5 cm]{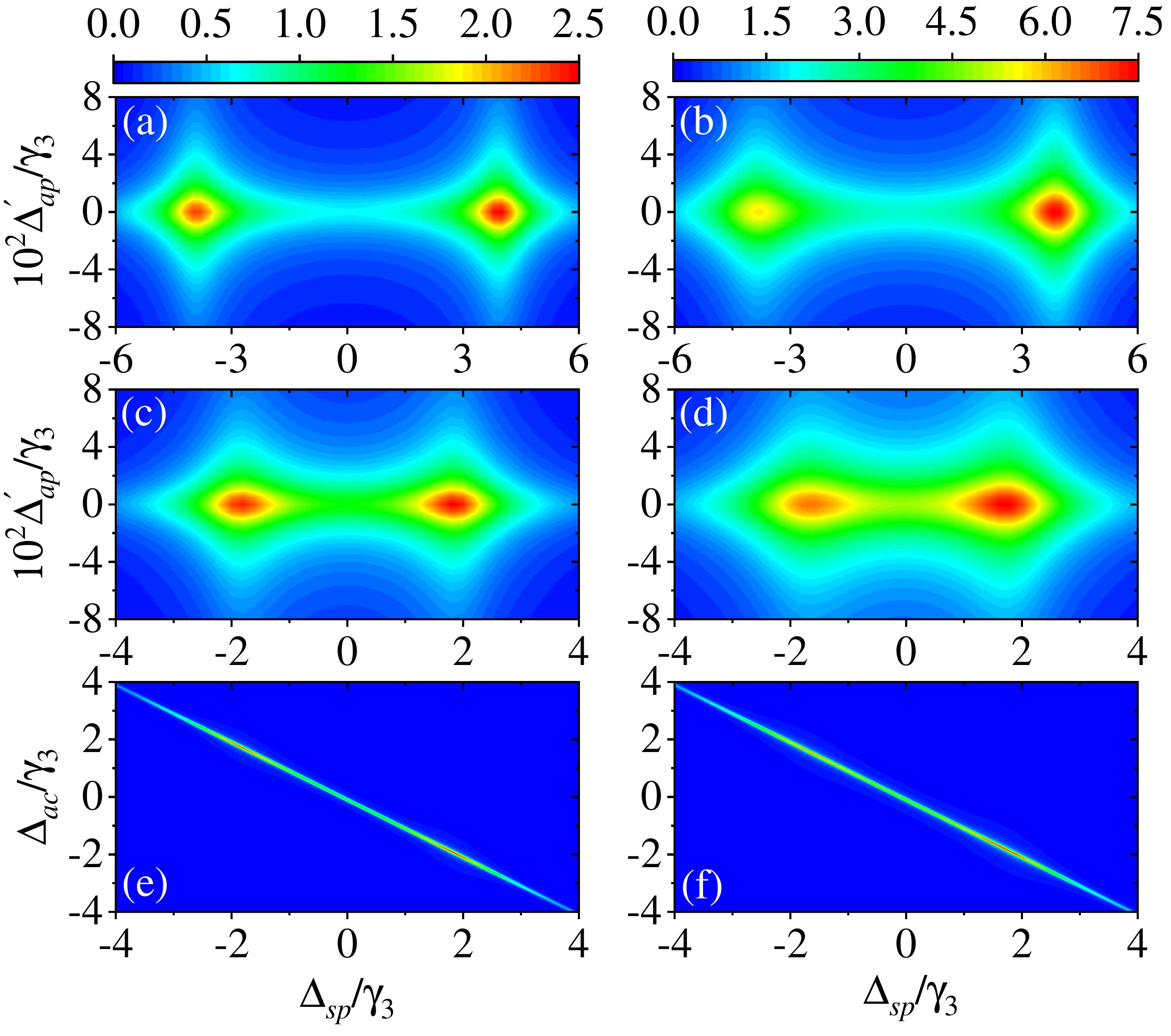}
\caption{Modulus of $10u_{1}^{sa}/u_{1}$ in units of $\gamma_3^{-1}$ against $\Delta_{sp}/\gamma_3$ and $\Delta'_{ap}/\gamma_3$ (a, b, c, d) or $\Delta_{sp}/\gamma_3$ and $\Delta_{ac}/\gamma_3$ (e, f) with $\Delta_{ac}=\Delta_{a}-\Delta_{c}$. Parameters are the same as in Fig.~\ref{fig2} except $\Gamma_{s}/\gamma_3=\Gamma_{a}/\gamma_3=0.1$ in (a, c, e) while $0.5$ in (b, d, f) as well as $\Omega_{c}/\gamma_3=4.0$ in (a, b) while $\Omega_{c}/\gamma_3=2.0$ in (c, d, e, f).} \label{fig3}
\end{figure}

\textit{Results and Discussion.} - We are now committed to verify the possibility of efficient biphoton generation via accurate numerical calculations. We start by examining single-atom spectral responses answering for the generation of Stokes-anti-Stokes biphotons. It is clear from Fig.~\ref{fig2} that both $u_{2}^{s}$ and $u_{4}^{s}$ exhibit a two-spike spectral structure with the spike separation determined just by $\Omega_{e}\simeq\sqrt{|\Omega_{c}|^2-\gamma_{4a}^2/4}\simeq\Omega_{c}$. This can be well understood by considering that $u_{2}^{s}$ and $u_{4}^{s}$ are different modification forms of the typical EIT spectrum described by 
\begin{equation}\label{Eq9}
\rho_{EIT}=\frac{-(\Delta_{sp}-i\gamma_{2})\Omega_p}{(\Delta_{sp}-i\gamma_2)(\Delta_{sp}-i\gamma_{4a})-|\Omega_c|^2},
\end{equation}
with $\Delta_{c}=0$. Its imaginary part exhibits two downward spikes as a pair of resonances at $\Delta_{sp}\simeq\pm\Omega_{c}$ but tends to zero at $\Delta_{sp}=0$ due to quantum destructive interference in the case of $\gamma_{2}\to0$. To be more specific, we have
\begin{equation}\label{Eq10}
\begin{split}
\frac{u_{2}^{s}}{u_{1}}&=\alpha\rho_{EIT}=\frac{\tilde{g}_s^*(\Delta_{sp}-i\gamma_{4a})\rho_{EIT}}{(\Delta_p+i\gamma_{3s})(\Delta_{sp}-i\gamma_2)},\\
\frac{u_{4}^{s}}{u_{1}}&=\beta\rho_{EIT}=\frac{\tilde{g}_s^*\Omega_c\rho_{EIT}}{(\Delta_p+i\gamma_{3s})(\Delta_{sp}-i\gamma_2)},
\end{split}
\end{equation}
with which it is easy to understand how $\alpha$ and $\beta$ modify the typical EIT spectrum to result in what are shown in Fig.~\ref{fig2}. For instance, we have $u_{2}^{s}/u_{1}\simeq-i\tilde{g}_s^*\Omega_{p}/\Delta_{p}\gamma_{4a}$ at $\Delta_{sp}\simeq\pm\Omega_{c}$ but $u_{2}^{s}/u_{1}\simeq -i\tilde{g}_s^*\Omega_{p}\gamma_{4a}/\Delta_{p}|\Omega_{c}|^2$ at $\Delta_{sp}=0$, which explains why $u_{2}^{s}/u_{1}$ exhibits an EIT-like two-spike response though its imaginary part at $\Delta_{sp}=0$ is not zero again and smaller in magnitude than that at $\Delta_{sp}\simeq\pm\Omega_{c}$ for $\Omega_{c}>\gamma_{4a}$. Moreover, we have $u_{4}^{s}/u_{1}\simeq\mp i\tilde{g}_s^*\Omega_{p}/\Delta_{p}\gamma_{4a}$ at $\Delta_{sp}\simeq\pm\Omega_{c}$ but $u_{4}^{s}/u_{1}\simeq\tilde{g}_s^*\Omega_{p}/\Delta_{p}\Omega_{c}$ at $\Delta_{sp}=0$, which explains why $u_{4}^{s}/u_{1}$ deviates largely from the EIT-type response with its imaginary part exhibiting a negative (positive) spike at $\Delta_{sp}\simeq\Omega_{c}$ ($\Delta_{sp}\simeq-\Omega_{c}$) and its real part being quite large in magnitude at $\Delta_{sp}=0$.

\begin{figure}[ptbh]
\includegraphics[width=8.5 cm]{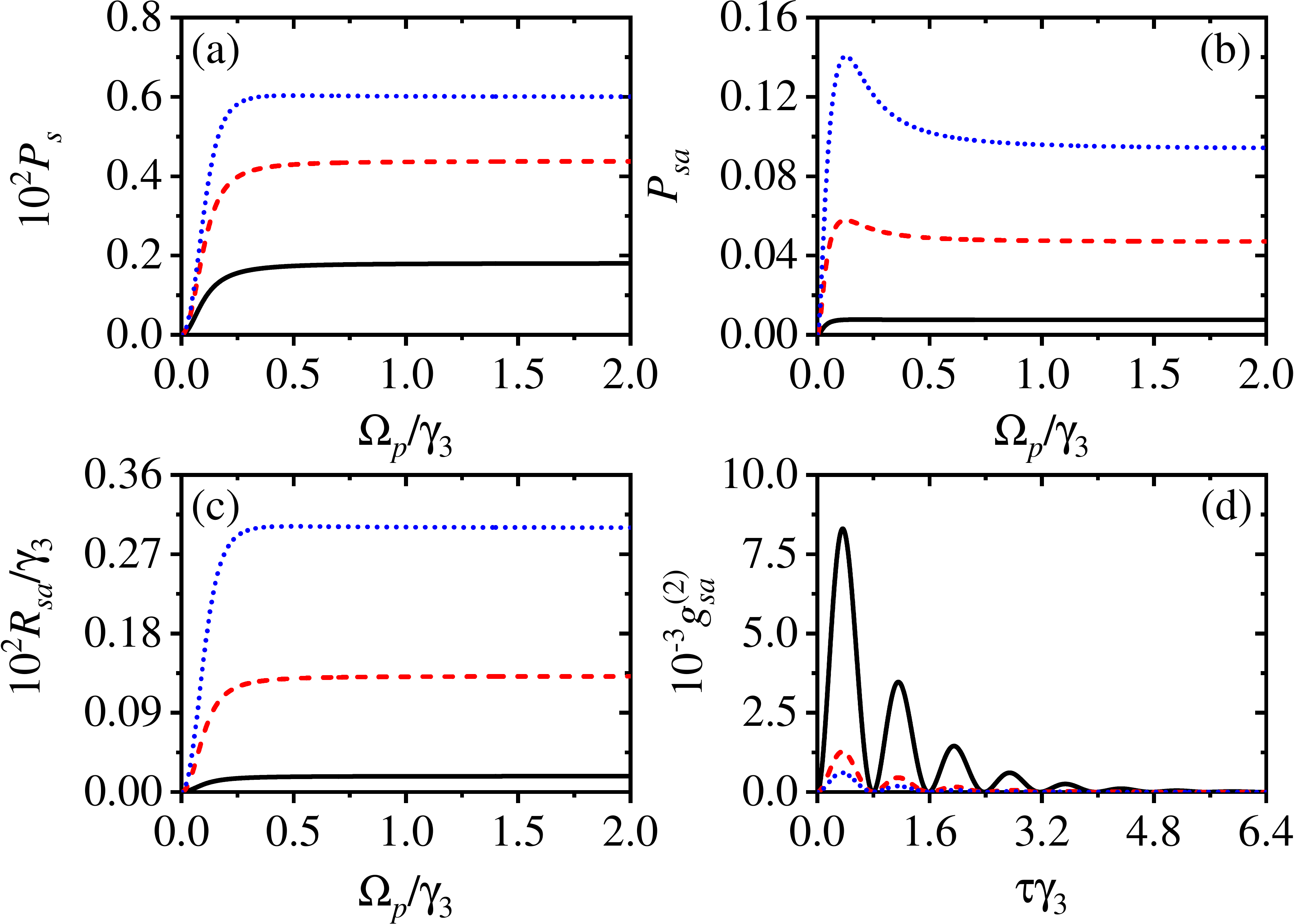}
\caption{$10^{2}P_{s}$ (a), $P_{sa}$ (b), and $10^{2}R_{sa}$ in units of $\gamma_{3}$ (c) against $\Omega_{p}/\gamma_{3}$ as well as $10^{-3}g_{sa}^{(2)}$ (d) against $\tau\gamma_{3}$ for $\Omega_{p}/\gamma_{3}=1.0$. Black-solid, red-dashed, and blue-dotted lines refer to $\Gamma_{s}/\gamma_{3}=\Gamma_{a}/\gamma_{3}=0.1$, $0.3$, and $0.5$, respectively. Other parameters are the same as in Fig.~\ref{fig2} except $\Omega_{c}/\gamma_{3}=4.0$.}\label{fig4} 
\end{figure}

{As to $u_{1}^{sa}/u_{1}$, its real and imaginary parts depend on $\Delta_{sp}$ in the same way as the imaginary and real parts of $u_{4}^{s}/u_{1}$, respectively, in the case of $\Delta_{ap}^{\prime}=0$, due to
\begin{equation}\label{Eq11}
\frac{u_{1}^{sa}}{u_{1}}=\frac{\tilde{g}_a^*}{\Delta_{ap}^{\prime}-i\gamma_{1}^{\prime}}\frac{u_{4}^{s}}{u_{1}}.
\end{equation}
We can also see from this equation that, for $\Delta_{ap}^{\prime}\ne0$, such a one-to-one correspondence is destroyed with both real and imaginary parts of $u_{1}^{sa}/u_{1}$ involving not only real part but also imaginary part of $u_{4}^{s}/u_{1}$ so that $u_{1}^{sa}/u_{1}$ must be asymmetric with respect to $\Delta_{sp}=0$. 

It is more instructive to examine in Fig.~\ref{fig3} the single-atom joint spectral response $|u_{1}^{sa}|$, which accounts for the generation of biphotons at two resonances $\Delta_{ap}^{\prime}=0$ and $\Delta_{sp}=\pm \mathrm{Re}(\Omega_{e})$. In particular, moving from the left-column panels to those on the right, we observe spectral intensities of increased asymmetries with growing values of $\Gamma_{s,a}$, as $\Omega_{e}$ turns complex. Meanwhile, the spectral intensities become slightly broader as $\gamma_{1}^{\prime}\propto\gamma_{3}+\Gamma_{s}$ increases with  $\Gamma_{s}$. Moving instead from the bottom panels to the top two, an increase of $\mathrm{Re}(\Omega_{e})$ results in more widely separated resonances controlled just by $\Omega_{c}$. Based on Figs.~\ref{fig3}(e) and \ref{fig3}(f), we further examine spectral correlations within each biphoton using the Schmidt number $K$. Rather than performing a full singular value decomposition~\cite{schmidt}, we approximate our $|u_{1}^{sa}|$ as a Gaussian ellipse, allowing $K$ to be computed analytically from the spectral amplitude ellipse geometry via two 1D fits instead of a complex 2D matrix decomposition~\cite{JMO}. We extract 1D profiles along the diagonal and anti-diagonal axes and estimate their peak widths to be  $\sigma_d \approx \gamma_1'$ and $\sigma_{ad} \approx 2\gamma_e$ from the analytical form of $u_1^{sa}$. The squared ratio $r = \sigma_d^2/\sigma_{ad}^2 \approx \gamma_1'^2/4\gamma_e^2 \approx 10^{-4}$~\cite{JMO} then yields $K=1/\sqrt{1-(1-r)^2/(1+r)^2}\simeq50$. This large Schmidt number due to a marked asymmetry in the widths of an elongated ellipse reflects strong spectral correlation (entanglement) between a pair of simultaneously generated Stokes and anti-Stokes photons.}

\begin{figure}[ptbh]
\includegraphics[width=8.5 cm]{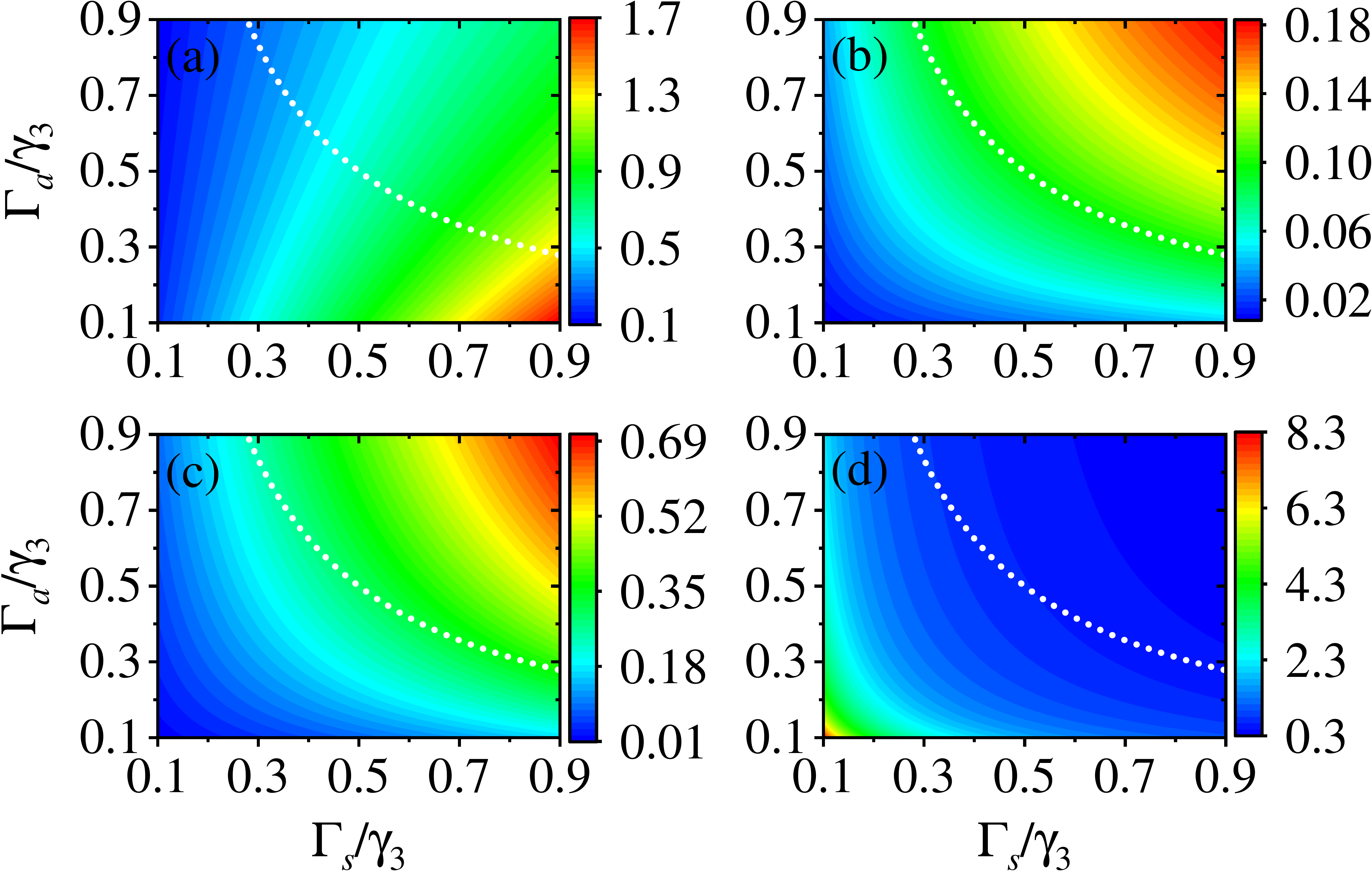}
\caption{$10^{2}P_{s}$ (a), $P_{sa}$ (b), $10^{2}R_{sa}$ in units of $\gamma_{3}$ (c), and $10^{-3}g_{sa}^{(2)}$ (d) at $\tau\gamma_{3}=0.36$ against $\Gamma_{s}/\gamma_{3}$ and $\Gamma_{a}/\gamma_{3}$. Other parameters are the same as in Fig.~\ref{fig2} except $\Omega_{c}/\gamma_{3}=4.0$.}\label{fig5} 
\end{figure}

Next, we transfer to evaluate the possibility of efficient biphoton generation by showing in Fig.~\ref{fig4} the dependence of $P_{s}$, $P_{sa}$, and $R_{sa}$ on $\Omega_{p}$ along with the dependence of $g_{sa}^{(2)}$ on $\tau$ by choosing three specific values of $\Gamma_{s}=\Gamma_{a}$. It is easy to find that $P_{s}$, $P_{sa}$, and $R_{sa}$ reach their respective saturation values for $\Omega_{p}/\gamma_{3}\gtrsim0.5$ and we always have $P_{sa}\gg P_{s}$, which is essential to ensure a larger proportion of the biphoton component $|\psi_{1}^{sa}\rangle$ relative to the single-photon components $|\psi_{2}^{s}\rangle$ and $|\psi_{4}^{s}\rangle$ in the eigenstate $|\psi\rangle$. It is also clear that, in principle, we can increase the saturation values of $P_{s}$, $P_{sa}$, and $R_{sa}$ by choosing relatively larger $\Gamma_{s}=\Gamma_{a}$, which cannot exceed $0.5\gamma$, however, to ensure $P_{sa}<0.1$ (\textit{i.e.}, $P_{sa}\ll|u_{1}|^2$) and hence avoid the simultaneous generation of two or more biphotons as already assumed in writing Eqs.~(\ref{Eq1}) and (\ref{Eq2}). In this regard, the biphoton generation rate $R_{sa}$ is typically of the order of $10^{-4}\gamma_{3}\sim10^{-3}\gamma_{3}$ for $\Gamma_{s,a}/\gamma_{3}>0.1$ and reaches its maximal value $R_{sa}^{\text{max}}\simeq3.0\times10^{-3}\gamma_{3}$ for $\Gamma_{s,a}/\gamma_{3}=0.5$. It is also important to note that the generated biphotons exhibit strongly nonclassical correlated features as manifested by the damped oscillations of $g_{sa}^{(2)}$ with an invariant period of $\pi/\Omega_{e}\simeq0.79/\gamma_{3}$ but gradually reducing maximal values of $8300$, $1270$, and $610$ for $\Gamma_{s,a}/\gamma_{3}=0.1$, $0.3$, and $0.5$, respectively, at $\tau\gamma_{3}\simeq0.36$. 

Finally, we examine in Fig.~\ref{fig5} the dependences of $P_{s}$, $P_{sa}$, $R_{sa}$, and $g_{sa}^{(2)}$ at $\tau\gamma_{3}\simeq0.36$ on $\Gamma_{s}$ and $\Gamma_{a}$ by considering that $\Gamma_{s}$ and $\Gamma_{a}$ are usually unequal due to $\mu_{32}\ne\mu_{41}$ for a specific real atom. It is clear that a smaller ratio of $\Gamma_{s}/\Gamma_{a}$ helps to well reduce the generation probability $P_{s}$ of a single Stokes photon as can be seen from Fig.~\ref{fig5}(a) and meanwhile remarkably enhance the nonclassical features of Stokes and anti-Stokes biphotons as can be seen from Fig.~\ref{fig5}(d). On the contrary, we find from Figs.~\ref{fig5}(b) and \ref{fig5}(c) that $P_{sa}$ and $R_{sa}$ exhibit more symmetric dependences on $\Gamma_{s}$ and $\Gamma_{a}$ so that they keep roughly invariant against a constant value of $\Gamma_{s}\Gamma_{a}$. Above results indicate that the optimal choice of relevant atomic levels should be featured by a larger $\Gamma_{a}$ ($\propto\mu_{41}^2$) and a smaller $\Gamma_{s}$ ($\propto\mu_{32}^2$) with their product restricted below a certain constant (\textit{e.g.}, $\gamma_{3}^2/4$) to ensure $P_{sa}\ll|u_{1}|^2$ as denoted by the white-dotted lines. Note also that $\Gamma_{s}$ and $\Gamma_{a}$ of ratio proportional to $\mu_{32}^2/\mu_{41}^2$ can be simultaneously increased by reducing the waveguide cross-sectional area $A$.

A possible experiment may be implemented by considering the Cs atom with its four levels $\vert1\rangle=\vert6S_{1/2},F=4,m_{F}=2\rangle$, $\vert2\rangle=\vert6S_{1/2},F=3,m_{F}=2\rangle$, $\vert3\rangle=\vert6P_{3/2},F=3,m_{F}=3\rangle$, and $\vert4\rangle=\vert6P_{3/2},F=4,m_{F}=3\rangle$ on the $D_{2}$ line~\cite{Alkali}. Thereby, we have $\gamma_{3,4}=2\pi \times2.61$ $\mathrm{MHz}$, $\mu_{14}=1.21\times10^{-29}$ Cm,  $\mu_{23}=1.16\times10^{-29}$ Cm, and $\lambda_{s,a}=2\pi c/\omega_{s,a}\simeq852\,\mathrm{nm}$ while the typically small $\gamma_{1,2}$ can be neglected. For $P_{sa}\le0.1$, $\Gamma_{a}/\gamma_{3}=1.09\Gamma_{s}/\gamma_{3}=0.1$ and $0.5$ require the waveguide to be featured by $\sqrt{A}=420\,nm$ and $188\,nm$, respectively, which are feasible with state-of-the-art nanofiber fabrication techniques. The corresponding biphoton generation rate $R_{sa}$ turns out to be $2.68\times10^{3}\,s^{-1}$ and $4.44\times10^{4}\,s^{-1}$, respectively, equivalent to (far exceeding) that can be attained with an atomic ensemble of $\mathcal{N}=10^{11}\, cm^{-3}$ while $\mathcal{L}=113\,\mu m$ and $566\,\mu m$, respectively in free space before (after) coupling into an on-chip setting as estimated by~\cite{review}
\begin{equation}\label{Eq12}
R_{sa}\simeq\frac{\pi^{2}\mathcal{L}^2\mathcal{N}^2\mu_{41}^2\mu_{23}^2}{8\epsilon_0^2\hbar^2\lambda_{s}\lambda_{as}\gamma_{41}}\frac{|\Omega_p|^2}{\Delta_{p}^{2}+\gamma_{31}^{2}}.
\end{equation}We may also consider the $^{87}$Rb atom with its four levels $\vert1\rangle=\vert5S_{1/2},F=1,m_{F}=0\rangle$, $\vert2\rangle=\vert5S_{1/2},F=2,m_{F}=0\rangle$, $\vert3\rangle=\vert5P_{1/2},F=1,m_{F}=1\rangle$, and $\vert4\rangle=\vert5P_{1/2},F=2,m_{F}=1\rangle$ on the $D_{1}$ line~\cite{Alkali}. Thereby, we have $\gamma_{3,4}=2\pi\times2.87$ $\mathrm{MHz}$, $\mu_{14}=1.27\times10^{-29}$ Cm, $\mu_{23}=0.73\times10^{-29}$ Cm, and $\lambda_{s,a}=2\pi c/\omega_{s,a}\simeq795\,\mathrm{nm}$. For $P_{sa}\le0.1$, $\Gamma_{a}/\gamma_{3}=3.0\Gamma_{s}/\gamma_{3}=0.2$ and $1.0$ require the waveguide to be featured by $\sqrt{A}=308\,nm$ corresponding to $R_{sa}=3.92\times10^{3}\,s^{-1}$ and $\sqrt{A}=138\,nm$ corresponding to $R_{sa}=5.25\times10^{4}\,s^{-1}$, respectively.  Though numerical estimations are made here for a real Cs or $^{87}$Rb atom, the underlying physical insights are universal indeed and can be generalized to other artificial-atom (\textit{e.g.}, semiconductor quantum dots and superconducting quantum circuits) coupled waveguide platforms~\cite{circuit,qdot}.

\bigskip    
\textit{Conclusions and Outlook.} - We have proposed a chip-scale platform for effectively generating high-quality photon pairs by implementing a single-atom SFWM within the waveguide QED framework. Leveraging enhanced atom-photon interactions afforded by transverse confinement of guided modes, the biphoton generation rate can exceed $10^4\,\text{s}^{-1}$ when a real four-level atom near a chiral waveguide is driven by a far-detuned pump and a resonant coupling field. This is comparable to that attained from a massive atomic ensemble in free space but requires no fiber coupling for practical applications, hence avoids substantial losses owing to severe mode mismatch. Note also that the interplay between atomic decay rates toward guided and non-guided radiation modes determine the optimal performance, which relies on low atomic excitation and moderate biphoton probability to well exclude single-photon and multi-biphoton generation events, respectively. This work then marks a significant extension of waveguide QED studies from usual linear single photon scattering to less explored nonlinear biphoton generation, providing a robust, low-loss, and highly integrable theoretical foundation for on-chip quantum light sources essential for scalable integrated quantum networks.

\textit{Acknowledgments.} - Supported by the National Natural Science Foundation of China (No.~62375047), Italian PNRR MUR (No.~PE0000023-NQSTI), I-PHOQS (Photonics and Quantum Sciences, PdGP/GePro 2024-2026), and the Fund for International Activities of the
University of Brescia.

\end{spacing}

\bibliography{refs.bib}
\end{document}